\documentstyle[aps,epsf]{revtex}
\def\tb{{\bf \overline{3}}}
\def\sx{{\bf 6}}
\def\det{{\rm det}}
\begin{document}
\begin{center}
{\Large \bf Why Color-Flavor Locking is Just like
Chiral Symmetry Breaking}
\end{center}
\vspace{0.3cm}

\begin{center}

{Robert D. Pisarski$^a$ and Dirk H. Rischke$^{b}$}
\end{center}
\vspace{0.3cm}
{\it $^a$ Department of Physics, Brookhaven National Laboratory,

Upton, New York 11973-5000, USA}\\
{\it $^b$ RIKEN BNL Research Center, Brookhaven National Laboratory,

Upton, New York 11973-5000, USA}
\vspace{0.5cm}

{\bf Abstract}
\vspace{0.5cm}

We review how a classification into representations of color and flavor
can be used to understand the possible patterns of symmetry breaking
for color superconductivity in dense quark matter.
In particular, we show how for three flavors, color-flavor locking
is {\it precisely\/} analogous to the usual pattern of chiral symmetry
breaking in the QCD vacuum.

\vspace{1.0cm}

While it has been known from the work of Bailin and Love 
that color superconductivity
can occur in cold, dense quark matter \cite{bl}, 
only recently has there been sustained interest,
following work of Alford, Rajagopal, and Wilczek, and of Rapp,
Sch\"afer, Shuryak, and Velkovsky 
\cite{general1,general2,arw1,abr,prlett,son,prscalar,qcdgap,nonbcs1,nonbcs2}.
Theories used to estimate the gap for color superconductivity include
Nambu--Jona-Lasinio (NJL) models and the renormalization group.
While Bailin and Love found tiny gaps,
$\sim 1$ MeV, current estimates of the gap are much larger,
$\sim 100$ MeV 
\cite{general1,general2,arw1,abr,prlett,son,prscalar,qcdgap,nonbcs1,nonbcs2}.

Color superconductivity is not like
``ordinary'' superconductivity,
such as in the theory of Bardeen, Cooper, and Schrieffer (BCS).
In the latter, the gap is an exponential in $1/g^2$.  This result
was also obtained by Bailin and Love
\cite{bl}, who assumed that, at zero temperature, 
static magnetic interactions are screened
by a (non-perturbative) magnetic mass. In QCD, however, 
Son \cite{son} and ourselves \cite{prlett} argued independently 
that nearly static magnetic interactions are not screened; 
if $g$ is the QCD coupling constant, collinear
emission of soft magnetic gluons turns the gap from an exponential
in $1/g^2$ into an exponential 
in $1/g$. For sufficiently small $g$,
$\exp(-1/g)$ is much larger than $\exp(-1/g^2)$.  This is one of the
reasons why the gap is now found to be two orders of magnitude larger
than the previous estimate by Bailin and Love.

The other reason is the following. When one explicitly
computes the magnitude of the gap for a spin-zero condensate, 
$\phi_0$, one finds
$$
\frac{\phi_0}{\mu} = 
b_0 \; \exp\left( - \frac{c_0}{g}\right) [ 1 + O(g)] \; , \;
$$
\begin{equation}
b_0 = 256 \; \pi^4 \left(\frac{2}{g^2 N_f}\right)^{5/2} b_0' \;\;\; ,\;\;\;
c_0 = \sqrt{\frac{6 N_c \pi^4}{N_c + 1}} \; .
\label{e0}
\end{equation}
This result is valid for $N_f$ flavors of massless quarks 
at a chemical potential $\mu$, coupled to
a $SU(N_c)$ gauge theory.
The value of $c_0$ and the prefactor of $1/g^5$ was computed first
by Son \cite{son}; the constant $256 \, \pi^4$ is due to
Sch\"afer and Wilczek \cite{nonbcs1} and ourselves \cite{nonbcs2}.
Actually, this number is only an {\it estimate\/} for the prefactor;
these are the factors from logarithms in the exponent 
which turn into a prefactor of $1/g^5$.
The uncertainty from other effects which also contribute
to the prefactor is represented by the number $b_0'$.

For a BCS gap, $\phi \sim \exp(-1/g^2)$, as a function of $g$
the gap is usually infintesimal until it turns on suddenly.
In contrast, because of the prefactor of $1/g^5$ in (\ref{e0}),
as $g$ increases from zero, 
$\phi_0/\mu$ increases, reaches a maximum,
and then {\it decreases\/} at strong $g$.  

(Incidentally, this is just like a semiclassical tunneling
probability, in which there are five zero modes.  Why the color
superconducting gap looks like a tunneling problem, with {\it five\/}
zero modes, {\it independent\/} of the number of colors or flavors, is
a complete mystery to us.)

Setting the undetermined constant $b_0'=1$, what is so exciting
is that because of the factor of $256 \, \pi^4$, this maximum value
of the gap is really quite large; for $N_f = 2$,
the maximum is $\phi_0/\mu \simeq 0.13$ and occurs at a coupling
of $g \simeq 4$, where $\alpha_s = g^2/4 \pi \sim 1$.
This large value of the gap
supports previous estimates found using NJL 
models \cite{general1,general2,arw1,abr}.  

(Notice also that
the gap {\it decreases\/} as $N_f$ {\it increases}, like $1/N_f^{5/2}$,
assuming~(!) that $b_0'$ does not depend significantly on $N_f$.
The implications of this for quark matter, where there are
up, down, and strange quarks, is intriguing.)

We also showed 
that the ratio of the transition temperature to the gap is
{\it twice\/} the same ratio in BCS, with
$T_c/\phi_0 \sim 1.13$ \cite{nonbcs2}.  This establishes
conclusively that color superconductivity is 
{\it not\/} like BCS.

In this talk we ignore the magnitude of the gap, to concentrate on
the possible patterns of symmetry breaking.  This has been analyzed
using three colors, and either two or three flavors, using NJL
models \cite{arw1,abr}.  In \cite{prlett} we developed a general
classification in terms of representations in color and flavor.
In this Proceeding we show that there is an exact analogy between
color-flavor locking for three flavors, and chiral symmetry breaking
in the QCD vacuum.  This was noted in our Letter \cite{prlett},
in refs.\ 19, 21, and 23, 
although perhaps the present relaxed discussion might
be more transparent than those cryptic comments.

Our analysis is motivated by the following
puzzle.  Scattering between two quarks occurs in one of two
possible ways, either 
anti-symmetric or symmetric in the two color indices for quarks
in the fundamental representation.
(The gluon interaction is flavor blind, but flavor will enter later
through Fermi statistics.)
In color $SU(3)_c$, for example,
the anti-symmetric representation is a color anti-triplet, $\tb$, and the 
symmetric is a color sextet, $\sx$.  Then in all models studied, either
NJL or with single-gluon exchange, the color anti-triplet 
channel is attractive,
and the color sextet channel is repulsive.  

(We comment that this need {\it not\/} be true for all values of the coupling.
Since QCD is asymptotically free, the color 
anti-triplet channel must be the
attractive channel for all densities above some value.  But in strong
coupling, which corresponds to intermediate densities, it is
certainly possible 
that the attractive channel is the color sextet.
The phenomenology of such an intermediate phase would be interesting
to explore.  We ignore this possibility for now, although 
it could be easily treated using our approach.)

For total spin $J=0$, the quark-quark condensate has the form
\begin{equation}
\phi^{i j}_{a b}(\Gamma) \; = \;
q^i_a(-\vec{p})^T \, C \, \Gamma \; q^j_b(\vec{p}) \; ,
\label{e1}
\end{equation}
$q$ is the quark field, $q^T$ the Dirac transpose, 
$C$ the charge conjugation matrix, and $\Gamma$ a 
Dirac matrix.  The $SU(3)_c$ color indices
for a quark in the fundamental representation are $i,j=1,2,3$;
the flavor indices are $a,b=1, \ldots , N_f$, for $N_f$ flavors
of massless quarks.  The Dirac matrix
$\Gamma$ classifies the chiral and helicity structure of the
condensate.  We don't worry about the four kinds of condensates 
\cite{prlett,prscalar}
which $\Gamma$ represents, since all that does is to add another index
to the condensate field.

For two flavors, the condensate found is of the form
\begin{equation}
\langle \phi^{i j}_{a b} \rangle = 
\epsilon^{i j k} \; \epsilon_{a b} \; \langle \phi^k \rangle\; .
\label{e2}
\end{equation}
This is anti-symmetric in the quark color indices, $i$ and $j$,
and so only involves
the color anti-triplet channel, as expected.
We are sloppy about indices: $\phi^k$ is an anti-triplet,
not a triplet.  It is also a flavor singlet.

For three flavors, one would also expect that the
condensate is again a color anti-triplet, 
\begin{equation}
\langle \phi^{i j}_{a b} \rangle = 
\epsilon^{i j k} \; \epsilon_{a b c} \; \langle \phi^k_c \rangle \; ,
\label{e3}
\end{equation}
and a flavor anti-triplet.

The puzzle is that 
this is {\it not\/} what is found.  Instead, 
the ``color-flavor locked'' condensate is of the form \cite{arw1,abr}
\begin{equation}
\langle \phi^{i j}_{a b} \rangle = 
\delta^i_a \; \delta^j_b \; \kappa_1
+ \delta^i_b \; \delta^j_a \; \kappa_2 \; .
\label{e4}
\end{equation}
This contains a piece which is anti-symmetric in the color
indices, and so is a color anti-triplet, but it also contains
a piece which is symmetric, and so is a color sextet.  
The color anti-triplet piece
is proportional to $\kappa_1 - \kappa_2$, the color sextet to
$\kappa_1 + \kappa_2$.  The color-flavor locked condensates 
\cite{arw1,abr} are all predominantly color triplet, but there is
always some color sextet piece, $0 \neq |\kappa_1 + \kappa_2| \ll |\kappa_1
- \kappa_2|$.
The quandry is then, why
for three flavors is there some condensation in a {\it repulsive\/} channel?
And why doesn't this happen for two flavors?

Crucial to the analysis is that for a spin-zero condensate of massless
particles, Fermi statistics requires that 
the condensate field is symmetric in the simultaneous interchange of
both color and flavor indices\cite{bl,prlett,prscalar}:
\begin{equation}
\phi^{i j}_{a b}(\Gamma) \; = \; + \, \phi^{j i}_{b a}(\Gamma) \; .
\label{e5}
\end{equation}
We assume that because of instantons, we can ignore the
difference between left- and right-handed chiral symmetries,
and speak only of a (global) $SU(N_f)_f$ symmetry.
Then we only need some trivial group theory: in $SU(2)$, 
${\bf 2}\times {\bf 2} = {\bf 1}_a + {\bf 3}_s$ in $SU(2)$, while
for $SU(3)$, ${\bf 3}\times {\bf 3} = \overline{{\bf 3}}_a + {\bf 6}_s$.
Here ``$a$'' and ``$s$'' denote anti-symmetric and symmetric
representations, respectively.  
By (\ref{e5}), then, we can have either representations in color
and flavor which are both anti-symmetric, or both symmetric.

For two flavors, under $(SU(3)_c,SU(2)_f)$ we can have a color anti-triplet,
flavor singlet, 
$(\overline{{\bf 3}}, {\bf 1})$, and a
color sextet, flavor triplet $({\bf 6}, {\bf 3})$.  The 
$(\overline{{\bf 3}}, {\bf 1})$ field is the $\phi^k$ of (\ref{e2}).

For three flavors, under $(SU(3)_c,SU(3)_f)$ we can have 
a color anti-triplet, flavor anti-triplet field
$(\overline{{\bf 3}}, {\overline {\bf 3}})$, as 
the $\phi^i_a$ field in (\ref{e3}), and 
a color sextet, flavor sextet field $({\bf 6}, {\bf 6})$.
The sextet field is denoted as $\chi^{i j}_{a b}$; the latter is symmetric
in each pair of indices, 
$\chi^{i j}_{a b} = + \chi^{j i}_{a b} =+ \chi^{i j}_{b a}$.

Effective lagrangians are constructed in the usual manner.
The kinetic terms include
\begin{equation}
{\cal L}_{0} = 
{\cal L}_{\rm gauge} + |D_\mu \phi|^2 + |D_\mu \chi|^2  \; .
\label{e60}
\end{equation}
Here the gauge action includes {\it both\/} the usual gauge
action for a nonabelian gauge theory, {\it and\/} the contribution
of ``Hard Dense Loops'' \cite{prlett,son,nonbcs1,nonbcs2}.
The kinetic terms for the condensate fields involve the
appropriate covariant derivative for that representation.
Thus if the covariant derivative for a triplet field is
$D_\mu = \partial_\mu - i g A_\mu$, for the anti-triplet
field $D_\mu \phi = (\partial_\mu + i g A_\mu)\phi$; the
covariant derivative for the sextet field is more involved.

Mass terms for the condensate fields are written to drive
condensation in the anti-triplet channel:
\begin{equation}
{\cal L}_{2} = 
- \mu_3^2 \; |\phi|^2 + \mu_6^2 \; |\chi|^2  \; .
\label{e6}
\end{equation}
If mixing between $\phi$ and $\chi$ could
be neglected, because of its negative mass squared term,
only $\phi$ condenses, and $\chi$ not.

(Condensation in the sextet channel is modeled simply
by taking a negative mass squared term for $\chi$, and a positive
one for $\phi$.)

There are many quartic terms which couple $\phi$ and $\chi$ together:
$$
{\cal L}_{4} = 
 \lambda_1 \; [{\rm tr}\, (\phi^\dagger \phi)]^2 
+ \lambda_2 \; {\rm tr}\, (\phi^\dagger \phi)^2 
+  \lambda_3 \; [{\rm tr}\, (\chi^\dagger \chi)]^2 
+  \lambda_4 \; {\rm tr}\, (\chi^\dagger \chi)^2 
$$
\begin{equation}
+ \lambda_5 \; {\rm tr}\,(\phi^\dagger \phi) \; {\rm tr}\,
(\chi^\dagger \chi) + \ldots
\label{e6b}
\end{equation}
plus other quartic terms, some of which mix $\phi$ and $\chi$.
We are opaque with respect to indices: 
${\rm tr}\, (\phi^\dagger \phi) = (\phi^i_a)^* \phi^i_a$, while
${\rm tr}\,(\phi^\dagger \phi)^2 = 
(\phi^i_a)^* \phi^j_a (\phi^j_b)^* \phi^i_b$.
Whatever the forms of the quartic terms, however, they are all
rather innocuous, since they are at least quadratic in the $\chi$ field.
Consequently, even 
if $\phi$ condenses, unless a coupling such as
$\lambda_5$ is large and negative,
it doesn't drive the condensation of 
$\chi$.  This is what is typical with two coupled fields: one field
may condense, but generally the other doesn't, so we can ignore the
field which doesn't.

For three flavors, however, there {\it are\/} operators which mix
$\phi$ and $\chi$ in a special way.  Consider
\begin{equation}
{\rm tr}\,( \phi \chi \phi) + {\rm h.c.} \equiv
\phi^i_a \; (\chi^{i j}_{a b}+ \chi^{i j}_{b a}) \; \phi^j_b + {\rm h.c.}
\label{e7}
\end{equation}
This is a cubic operator which is invariant 
under both $SU(3)_c$ color and $SU(3)_f$ flavor
rotations.  Our sloppiness in notation obscures this point:
the indices on $\phi$ are anti-triplets,
and so can be contracted with the triplet indices on $\chi$.

Nevertheless, such an operator does not appear in an effective lagrangian,
because it is not invariant under global $U(1)$ rotations of baryon
number.  This is easily corrected, however, by multiplying
by the operator $\det(\phi)^*$; this is also invariant under 
$SU(3)_c$ color and $SU(3)_f$ flavor, but precisely soaks up
the requisite factors for baryon number.  Thus there is no symmetry
reason to prevent the following term from appearing in the
effective lagrangian:
\begin{equation}
{\cal L}_6 = \lambda_6 \; 
\det(\phi)^* \; {\rm tr}\,(\phi \chi \phi) + {\rm h.c.}
\label{e8}
\end{equation}
This is a six-point operator, and so in
the sense of the renormalization group, is marginal
relative to ${\cal L}_4$, and doesn't affect the critical behavior.

It does, however, affect the {\it pattern\/} 
of condensation.  In particular, assume
that $\phi^i_a$ condenses in such a manner that $\det(\phi) \neq 0$.
This is true for color-flavor locking, where 
\begin{equation}
\langle \phi^i_a \rangle = \delta^i_a \; \phi_0 \; .
\label{e9}
\end{equation}
Then our funny operator becomes
\begin{equation}
{\cal L}_6 = \lambda_6 \, \phi_0^5 \; (\chi^{i j}_{i j} 
+ \chi^{i j}_{j i}) + {\rm h.c.}\; .
\label{e10}
\end{equation}
Thus when $\phi$ condenses, ${\cal L}_6$ generates a term
{\it linear\/} in $\chi$.  The special thing about an operator
linear in a field is that even if the mass squared term is positive,
a nonzero vacuum expectation value is always generated by a linear
term.  Thus even when $\mu_6^2 > 0$, 
$\chi$ acquires a nonzero expectation value
from its mixing with $\phi$,
\begin{equation}
\langle \chi^{i j}_{a b} \rangle \sim \lambda_6 \frac{\phi_0^5}{\mu_6^2}\;
(\delta^i_a \delta^j_b + \delta^i_b \delta^j_a)\; .
\label{e11}
\end{equation}
There are of course many other terms which couple six fields together.
None of these other terms, however, produce a term linear in $\chi$
when $\phi$ condenses.

(One can also write down a six-point term like that in
(\ref{e8}), except with $\det(\phi)$ replaced by $\det(\chi)$.  Even
with such a term, however, if for intermediate densities condensation
is for the sextet instead of the anti-triplet field, 
by switching the sign of the mass terms in 
(\ref{e6}), $\chi$ condensation does {\it not\/} drive $\phi$ condensation,
since the six-point term like that in (\ref{e8}) is still quadratic
in $\phi$, not linear.)

It is now easy to understand why we don't have to worry about
this kind of mixing for two flavors.  When $N_f=2$,
the $(\tb,{\bf 1})$ is $\phi^i$, 
and the $(\sx,{\bf 3})$ field $\chi^{i j}_{a b}$.
There is simply no cubic coupling possible between two $\phi^i$'s and one
$\chi^{i j}_{a b}$; there is also no $\det(\phi)$ term.  Thus when
the color anti-triplet $\phi$ condenses, the 
color sextet $\chi$ does {\it not}.
This is what detailed calculations in NJL models 
find \cite{general1,general2}.

The physically realistic case is $2+1$ flavors, with the up and down
quarks essentially degenerate in mass, but the strange quark 
with a large mass.  One can then show that $2+1$ flavors is
analogous to three flavors, although the proliferation of fields
becomes a little dizzying.  
For the condensates of up quarks with
down quarks, denote the color anti-triplet field as
$\phi^i$, and the color sextet field as $\chi^{i j}_{a b}$.
For the condensates of up or down quarks with strange quarks,
we write the color anti-triplet field as $\widetilde{\phi}^i_a$, and the
color sextet field as $\widetilde{\chi}^{i j}_a$.  Here the
flavor index is only for up and down, $a,b=1,2$.
Then there is an operator analogous to $\det(\phi)$; it is
\begin{equation}
\det(\phi) \rightarrow \epsilon^{i j k} \epsilon^{a b}
\phi^i \widetilde{\phi}^j_a \widetilde{\phi}^k_b \; .
\label{e12}
\end{equation}
The existence of this operator is interesting in its own right.
For three flavors, Sch\"afer and Wilczek \cite{abr} pointed out
that $\det(\phi)$ can be used as a global order parameter to distinguish
different phases.  The operator of (\ref{e12}) is the analogy
for $2+1$ flavors.

For $2+1$ flavors, the operator similar to (\ref{e8}) is
\begin{equation}
\widetilde{\cal L}_6 = 
\epsilon^{i j k} \epsilon^{a b}
\phi^i \widetilde{\phi}^j_a \widetilde{\phi}^k_b \; 
\left( \lambda_6' \; \widetilde{\phi}^l_c \;
\chi^{lm}_{cd} \; \widetilde{\phi}^m_d  
+ \lambda_6^{''} \; \phi^l \; 
\widetilde{\chi}^{lm}_c \; \widetilde{\phi}^m_c  \right) + \ldots
\label{e20}
\end{equation}
Because of this operator, when {\it both\/}
$\phi^i$ and $\widetilde{\phi}^i_a$ condense, so
will the two color sextet fields, $\chi^{i j}_{a b}$ and
$\widetilde{\chi}^{i j}_a$.

This conclusion agrees with all previous work done using NJL models
\cite{abr}.  It is not apparent from their analysis, which did
not use our color and flavor decomposition, but it does explain
why, in terms of (\ref{e4}), $\kappa_1 + \kappa_2 \neq 0$.
It also agrees with 
Alford, Berges, and Rajagopal \cite{abr}, who find that 
for $2+1$ flavors, the color sextet
field condenses {\it only\/} along the direction in which the color
anti-triplet field condenses.  This is what one expects 
from the detailed form of the cubic coupling in (\ref{e20}).
One doesn't see the color sextet field 
in weak coupling \cite{son,qcdgap,nonbcs1,nonbcs2}, since it 
is of fifth order in the color anti-triplet condensate. 

Now we are ready to understand the analogy to chiral symmetry
breaking.
In our analysis, the admixture of a color symmetric piece in 
the condensate of (\ref{e4}) is really a red herring; what matters
is the condensation of $\phi^i_a$ in (\ref{e9}).  
Once we recognize that we can basically forget about the color sextet
field, and concentrate on the color anti-triplet, 
we see that the pattern of symmetry
breaking is most familiar.  Consider
not $SU(3)_c$ color and $SU(3)_f$ flavor, but chiral symmetry
breaking, where a global flavor symmetry of 
$SU(3)_\ell \times SU(3)_r$ breaks to $SU(3)_f$.
The order parameter is again a $3 \times 3$ matrix,
a $(\overline{{\bf 3}},{\bf 3})$ under $(SU(3)_\ell,SU(3)_r)$ flavor:
\begin{equation}
U^{a b} = \; \overline{q}^i_{\ell,a} \; q^i_{r,b}
\end{equation}
so $a$ and $b$ are triplet indices for the flavor symmetries of $SU(3)_\ell$
and $SU(3)_r$, respectively.  The flavor matrix $U^{a b}$ is of
course a color singlet.

(While it does not happen in the vacuum,
in a phase in which color superconductivity
occurs, if chiral symmetry is broken in the color singlet channel,
it may well condense in a color octet channel as well.  To be precise,
one should classify the chiral symmetry with respect to the unbroken
color/flavor group.)

Then in QCD,
the observed pattern of chiral symmetry breaking is {\it identical\/} to
that of (\ref{e9}).  For chiral symmetry breaking, 
the condensate like that of (\ref{e9}) is
\begin{equation}
\langle U^{a b} \rangle = \delta^{a b} \, U_0 \; ,
\end{equation}
and corresponds to {\it all\/} quark flavors developing the {\it same\/}
constituent quark mass.
It is crucial to drop the color sextet piece from the condensate
in order to understand this analogy; it is not at all obvious from
(\ref{e4}).

This is not the only way in which the symmetry may break;
instead, only {\it one\/} flavor  of quark might develop a constituent quark
mass, $\langle U^{a b} \rangle \sim 
\delta^a_1 \delta^b_1 $.  
(These are the only two possibilities \cite{prlett}.)
For chiral symmetry breaking in QCD, 
nature obviously favors a common constituent quark mass for all 
flavors over just one flavor developing a quark mass.
This was proven by 
Coleman and Witten to be true in the  
limit of a large number of colors, $N_c \rightarrow \infty$ \cite{coleman}.

For color superconductivity, the pattern of symmetry breaking
analogous to one flavor of quark developing a constituent quark mass is
\begin{equation}
\langle \phi^i_a \rangle = \; \delta^{i 1} \; \delta_{a 1} \; \phi_0' \; .
\label{e13}
\end{equation}
It is easy to argue why the symmetry breaking 
in (\ref{e9}) is favored over that in (\ref{e13}):
(\ref{e9}) ensures that all colors develop a gap, while
(\ref{e13}) leaves most colors ungapped. Gapping all colors is
energetically favorable.

We conclude by making a simple point about the effects of terms
in the effective lagrangian for nonzero current quark masses.  For
chiral symmetry, these terms are linear in the $U$ field,
\begin{equation}
{\cal L}_{{ \rm quark \; mass}} \sim {\rm tr}\,[ m_q (U + U^\dagger) ] \; .
\label{e14}
\end{equation}
Consequently, the pseudo-Goldstone bosons, such as pions, {\it etc.},
have a mass squared which is {\it linear\/} in the quark mass,
$m^2_\pi \sim m_q$.

A term linear in ${\rm tr}\,(U)$ is possible because we only need to preserve
the flavor symmetry of $SU(N_f)_f$.  In contrast, mass terms are
very different for color superconductivity, since we still have
to preserve the invariance under color and flavor.  For example,
if we were to add a term for the strange quark mass to a three
flavor lagrangian, it would be {\it quadratic\/} in the strange quark
mass:
\begin{equation}
{\cal L}_{{ \rm quark \; mass}} \sim m_{s}^2 \; 
(\phi^i_s)^* \phi^i_s \; ,
\label{e15}
\end{equation}
where $m_s$ is the strange quark mass, and ``$s$'' denotes the flavor
direction for the strange quark.  This is rather heavy:
this pseudo-Goldstone boson has a mass like the strange
quark mass, $\sim 100$ MeV.  

A more interesting example, which we do not have space to explain, arises when
one considers the ``instanton-free'' region which arises at very
high densities.  In that case we do effectively restore the
full flavor symmetry, and neglecting instanton and mass effects,
parity is spontaneously broken \cite{prlett,prconf}.  Then we
must introduce left- and right-handed condensate fields, and like
(\ref{e15}), the mass term is
\begin{equation}
{\cal L}_{{ \rm quark \; mass}} \sim (m_{u} + m_{d})^2 
|\phi^i_{\ell,a} + \phi^i_{r,a}|^2\; ,
\label{e16}
\end{equation}
where $m_{u}$ and $m_{d}$ are the up and down quark masses.
As in (\ref{e15}), the pseudo-Goldstone boson for the breaking
of parity has a mass which is {\it linear\/} in the quark mass.
But now we are dealing with the up and down quark masses, which
are {\it very\/} light, $\sim 5$ or $10$ MeV; thus the mass
for the pseudo-Goldstone boson is also of this order.  

In heavy-ion collisions, this implies that if one reaches a phase in
which cool, dense quark matter is produced in an instanton-free regime,
then the time scale for a parity violating condensate to relax is
not $\sim 1$ fm, as it is with (say) disoriented chiral condensates,
but a factor of ten or twenty larger, $\sim 10-20$ fm/c.  This might
produce observable signals, if one is careful to trigger on collisions
in which the produced quark matter is not hot, but cool.

\end{document}